\documentclass[superscriptaddress,showpacs,showkeys,preprintnumbers,nofootinbib,amsmath,amssymb,aps,onecolumn,prd,floatfix,11pt]{revtex4-2}

\begin{document}

\title{New Integrable Chiral Cosmological Models with Two Scalar Fields}

\author{Vsevolod~R.~Ivanov,}\email{vsvd.ivanov@gmail.com}
\affiliation{Physics Department, Lomonosov Moscow State University\\
Leninskie Gory~1, 119991, Moscow, Russia}
\author{Sergey~Yu.~Vernov}\email{svernov@theory.sinp.msu.ru}
\affiliation{Skobeltsyn Institute of Nuclear Physics, Lomonosov Moscow State University, Leninskie Gory 1, 119991, Moscow, Russia}

\begin{abstract}
We construct integrable chiral cosmological models  with two scalar fields and potentials represented in terms of hyperbolic functions. Using the conformal transformation of the metric and the corresponding models with induced gravity terms, we obtain the general solutions in the spatially flat, open and closed Friedmann universes and the corresponding integrals of motion.
The obtained general solutions can be written in terms of the Jacobi elliptic functions of the conformal time.
\end{abstract}


% 04.50.Kd 	Modified theories of gravity
% 04.20.Jb 	Exact solutions
% 98.80.-k  Cosmology


\maketitle

\label{sec:intro}
\section{Introduction}

Many cosmological models, which describe the global evolution of the Universe, include scalar fields. To describe only one epoch of the Universe evolution, single-field models can be used. Using two-field models, one can explore multiple epochs or multiple physical phenomena. Models with a single scalar field nonminimally coupled to gravity can be transformed to models with a minimally coupled scalar field with a canonical kinetic term by metric and scalar field transformations. On the other hand, it is not possible to transform a model with two scalar fields nonminimally coupled to gravity to a model with two minimally coupled scalar fields and a standard kinetic part of the Lagrangian in the most general case~\cite{Kaiser:2010ps}. After the metric transformation, one obtains a general relativity model with nonstandard kinetic terms of scalar fields, a so-called chiral cosmological model (CCM)~\cite{Chervon:1995jx}.

In the Friedmann-Lema\^{i}tre-Robert\-son-Walker (FLRW) metric, the Einstein's equations reduce to a system of ordinary differential equations.
The problem of searching for integrable cosmological models with scalar fields is being actively investigated, essentially in the case of single-field cosmological models, but such models have been found only for a few specific forms of the scalar field potential~\cite{Salopek:1990jq,Russo:2004ym,Elizalde:2004mq,Maciejewski:2008hj,Bars:2010zh,Andrianov:2011fg,Bars:2012mt,Fre:2013vza,Fre:2013tya,
Kamenshchik:2013dga,Boisseau:2015hqa,Kamenshchik:2015cla,Sokolov:2016zjl,Kamenshchik:2017ojc,Paliathanasis:2018vru,Capozziello:2018gms,Fermi:2020yfd}.

A two-field CCM with a constant potential is trivially integrated not only in the spatially flat FLRW metric, but also in the Bianchi I metric~\cite{Ivanov:2021ily,Ivanov:2023uvh}. At the same time, the search for integrable cosmological models with multiple scalar fields and nonconstant potentials is an extremely complicated task. We know only a few examples of integrable cosmological models with two or more fields~\cite{Chimento:1998ju,Christodoulidis:2018msl,Paliathanasis:2018vru,Giacomini:2021xsx,Christodoulidis:2021vye,Russo:2023nir}.
The goal of this paper is to construct new integrable CCMs with two scalar fields and potentials written in terms of hyperbolic functions.

There are a few methods to integrate evolution equations in the FLRW metric. The integrability can be proven by using a suitable parametric time~\cite{Fre:2013vza}, but it is not immediately clear as to how to find this parametric time.
The superpotential method, which is also known as the Hamilton-Jacobi equation approach, is useful to integrate not only single-field models with the exponential potential~\cite{Salopek:1990jq}, but also two-field
models with the exponential potential~\cite{Russo:2023nir}. Note that this method is actively used to study inflation and to construct cosmological models with particular exact solutions~\cite{Muslimov:1990be,Salopek:1990jq,Liddle:1994dx,Kinney:1997ne,
Arefeva:2004odl,Bazeia:2005tj,Arefeva:2005mka,Skenderis:2006rr,Vernov:2006dm,Townsend:2007aw,Chervon:2008zz,Arefeva:2009tkq,Kamenshchik:2012pw,Harko:2013gha,Binetruy:2014zya,Vernov:2019ubo,Chervon:2019nwq,Fomin:2020caa,Russo:2023nir,Kamenshchik:2024kay}.

Another method for constructing single-field integrable models is based on the conformal metric transformation and the construction of the corresponding model in the Jordan frame~\cite{Bars:2010zh,Kamenshchik:2013dga,Kamenshchik:2015cla,Kamenshchik:2016atu}. In Ref.~\cite{Bars:2010zh}, the authors construct two-field models with one ordinary scalar field, one phantom field, and induced gravity terms to prove the integrability of a single-field cosmological model with a potential written in terms of hyperbolic functions. Note that integrability of this model has been proven in the closed and open Friedmann universes as well~\cite{Bars:2012mt}. In Refs.~\cite{Bars:2010zh,Bars:2011mh,Bars:2011aa}, the proposed cosmological model has been intensively studied in the context of the early Universe evolution, with special emphasis on inflation and the possibility of crossing the big bang-big crunch singularity.
In Ref.~\cite{Kamenshchik:2015cla}, the authors have proven integrability of the same model in a simpler way by using a single-field integrable model with nonminimal coupling constructed in Ref.~\cite{Boisseau:2015hqa}.
Note that this  model belongs to a wide class of integrable models with minimally coupled scalar fields and potentials in terms of hyperbolic functions. Their integrability have been proven by using a suitable parametric time in Ref.~\cite{Fre:2013vza} (see also Ref.~\cite{Kamenshchik:2015cla}).

To construct a new integrable CCM with two fields, we start from models with two nonminimally coupled scalar fields. The single-field integrable model proposed in Ref.~\cite{Boisseau:2015hqa} has an interesting feature: the Ricci scalar is an integral of motion. Recently, $N$-field cosmological models with the same property have been found and their integrability in the spatially flat FLRW metric have been proven~\cite{Ivanov:2024yjr}.

In this paper, we find general solutions of evolution equations in the Friedmann universe with arbitrary spatial curvature for a few such two-field integrable models with nonminimal coupling.
We show that in the conformal time, the system of the evolution equations can be transformed to a Hamiltonian system of two equations. For a few two-field polynomial potentials, we integrate this system and get analytic expressions of the general solutions in terms of the Jacobi elliptic functions. After finding such solutions, we obtain two-field chiral cosmological models in the Einstein frame by the conformal transformation of the metric.
The obtained integrable CCMs have potentials represented in terms of hyperbolic functions.

The structure of the paper is as follows. In Section~II, we consider two-field models with induced gravity terms. In Section~III, we construct the corresponding CCMs by the conformal transformation of the metric and the introduction of new fields. In Section~IV, we find solutions in the FLRW metric for the model with the nonminimal coupling. Examples of integrated CCMs and their general solutions in the conformal time are given in Section~V.
The results are summarized in Section~VI.

\section{Two-field model with nonminimal coupling}

Let us consider models with two nonminimally coupled scalar fields, described by an action
\begin{equation}
\label{actionJF}
S=\!\int\! d^4 x \sqrt{-g} \left[U(\sigma^1, \sigma^2) R - \frac{1}{2}g^{\mu \nu}\left(C_1\partial_\mu\sigma^1\partial_\nu\sigma^1+C_2\partial_\mu\sigma^2\partial_\nu\sigma^2\right) - V(\sigma^1, \sigma^2)\right]\!,
\end{equation}
where the functions $U$ and $V$ are differentiable, $R$ is the Ricci scalar,
and $C_A$ are constants. Positive values of $C_A$ correspond to the case of standard scalar fields.

By varying the action (\ref{actionJF}), we obtain evolution equations
\begin{equation}
\label{equRmunu}
U\left(R_{\mu \nu} - \frac12 g_{\mu \nu} R\right) = \nabla_\mu \nabla_\nu U - g_{\mu \nu} \Box U + \frac12 T_{\mu \nu},
\end{equation}
where
\begin{equation*}
T_{\mu \nu} =C_1\partial_\mu\sigma^1\partial_\nu\sigma^1+C_2\partial_\mu\sigma^2\partial_\nu\sigma^2-g_{\mu \nu} \left(\frac{1}{2}g^{\alpha \beta}\left(C_1\partial_\alpha\sigma^1\partial_\beta\sigma^1+C_2\partial_\alpha\sigma^2\partial_\beta\sigma^2\right) + V\right)\,,
\end{equation*}
and the field equations are
\begin{equation}
\label{equfield}
C_A\Box \sigma^A - V_{,\sigma^A} + R U_{,\sigma^A} = 0,
\end{equation}
$A=1,2$; $\, F_{,\sigma^A} \equiv \frac{\partial F}{\partial\sigma^A}$  for any function $F(\sigma^1,\sigma^2)$, and there is  no summation in $A$.

From Eq.~(\ref{equRmunu}), it follows that
\begin{equation}
\label{equR}
    3\Box U -UR=\frac12 g^{\mu\nu}T_{\mu\nu}={ } - \frac12 g^{\alpha \beta}\sum\limits_{A=1}^2C_A\partial_\alpha\sigma^A\partial_\beta\sigma^A - 2V.
\end{equation}

Our goal is to find such functions $U$ and $V$ that Eq.~\eqref{equR} has an integral of motion.
Using the field equations (\ref{equfield}), we get
\begin{equation}
\label{boxU}
\Box U=\sum\limits_{A=1}^2\frac{1}{C_A}\left(U_{,\sigma^A}V_{,\sigma^A}-R\left(U_{,\sigma^A}\right)^2\right)+\sum\limits_{A,B=1}^2U_{,\sigma^A\sigma^B}g^{\alpha\beta}\partial_\alpha\sigma^A\partial_\beta\sigma^B.
\end{equation}

Substituting Eq.~(\ref{boxU}) into Eq.~\eqref{equR}, we obtain
\begin{equation}
\label{equRUf}
 2V+\sum\limits_{A=1}^2 \frac{3}{C_A}U_{,\sigma^A}V_{,\sigma^A}=R\left(U+\sum\limits_{A=1}^2\frac{3}{C_A}\left[U_{,{\sigma^A}}\right]^2\right)+\frac{g^{\alpha\beta}}{2}\sum_{A,B=1}^2\left[6 U_{,\sigma^A\sigma^B}+C_A\delta_{AB}\right]\partial_\alpha \sigma^A \partial_\beta \sigma^B.
\end{equation}

To solve Eq.~(\ref{equRUf}), we choose such a function $U(\sigma^1,\sigma^2)$ that
\begin{equation}
\label{equU}
6 U_{,\sigma^A\sigma^B}+C_A\delta_{AB}=0,
\end{equation}
for all values of $A$ and $B$,
and
\begin{equation}
\label{equU0}
    U+\sum\limits_{A=1}^2\frac{3}{C_A}\left[U_{,{\sigma^A}}\right]^2=U_0,
\end{equation}
where $U_0$ is a constant.

The solution of Eqs.~(\ref{equU}) and (\ref{equU0}) is
\begin{equation}
\label{U_choice}
U(\sigma^1, \sigma^2)=U_0-\frac{C_1}{12}\left(\sigma^1-\sigma^1_0\right)^2-\frac{C_2}{12}\left(\sigma^2-\sigma^2_0\right)^2,
\end{equation}
where $\sigma^A_0$ are constants.

With such a function $U$, Eq.~(\ref{equRUf}) takes the form
\begin{equation}
\label{diffequV}
 2V-\frac{1}{2}\sum\limits_{A=1}^2 \sigma^AV_{,\sigma^A}=RU_0.
\end{equation}

If $R$ is a constant, $R=R_0$, then Eq.~(\ref{diffequV}) is a linear first-order partial differential equation with $V(\sigma^1,\sigma^2)$ as an unknown function. Equation~(\ref{diffequV}) has the following general solution:
\begin{equation}
\label{V_choice}
V = \frac{R_0 U_0}{2} + \left(\sigma^1 - \sigma_0^1\right)^4 f\left(\frac{\sigma^2 - \sigma_0^2}{\sigma^1 - \sigma_0^1}\right),
\end{equation}
where $f$ is an arbitrary differentiable function.

Note that for models with functions $U$ and $V$ given by Eqs.~(\ref{U_choice}) and (\ref{V_choice}), the Ricci scalar $R$ is the integral of motion in an arbitrary metric.
This result can be generalized on models with an arbitrary number of scalar fields~\cite{Ivanov:2024yjr}.

We consider the case of $U_0>0$, when the function $U>0$ for some values of $\sigma^A$. The case of $U_0<0$ corresponds to antigravity. If $U_0=0$, then Eq.~(\ref{diffequV}) restricts the potential without the assumption that $R$ is a constant. It means that nontrivial solutions of the Friedmann equations can exist only for potential (\ref{V_choice}) with $U_0=0$. In the case of single-field models, an analogous restriction on the potential has been obtained in Ref.~\cite{Arefeva:2012sqa}.

\section{The corresponding models in the Einstein frame}
Now we construct models in the Einstein frame that correspond to the previously chosen functions $U$ and $V$.
To get a model with two standard scalar fields, we take all $C_A=1$. Also, we can take $\sigma^A_0=0$ without loss of generality.

By doing a conformal transformation of the initial metric $g_{\mu \nu}$ of the form
\begin{equation}
g^{(\mathrm{E})}_{\mu \nu} = \frac{2 U}{M^2_\mathrm{Pl}} g_{\mu \nu},
\end{equation}
one arrives at the following action in the Einstein frame:
\begin{equation}
\label{SE}
S_\mathrm{E} = \int d^4 x \sqrt{-g^{(\mathrm{E})}} \left(\frac{M^2_\mathrm{Pl}}{2} R_\mathrm{E} - \frac{g^{(\mathrm{E})\mu \nu}}{2} \sum\limits_{A,B=1}^2 K_{AB}\partial_\mu \sigma^A \partial_\nu \sigma^B - V_\mathrm{E}\right),
\end{equation}
where $M_{\mathrm{Pl}}$ is the reduced Planck mass,
\begin{equation}
\label{kinetic_mat_gen_form}
K_{AB} = \frac{M^2_\mathrm{Pl}}{2 U} \left(\delta_{AB} + \frac{3}{U}  U_{,\sigma^A} U_{,\sigma^B}\right),
\end{equation}
\begin{equation}
\label{VE}
V_\mathrm{E} = \frac{M^4_\mathrm{Pl}}{4 U^2} V.
\end{equation}

For the chosen function $U(\sigma^1,\sigma^2)$ the matrix $K_{AB}$ is not diagonal:
\begin{equation}
\begin{aligned}
K_{11} =& \frac{M^2_\mathrm{Pl}}{2 U^2} \left(U + \frac{1}{12}\left(\sigma^1\right)^2\right),\\
K_{12} =& K_{21} = \frac{M^2_\mathrm{Pl}}{24 U^2} \sigma^1 \sigma^2,\\
K_{22} =& \frac{M^2_\mathrm{Pl}}{2 U^2} \left(U + \frac{1}{12}\left(\sigma^2\right)^2\right).
\end{aligned}
\end{equation}

It is possible to diagonalize the matrix $K_{AB}$ by introducing new fields $\chi^1$ and $\chi^2$. In general, one has
\begin{equation*}
K_{CD} \partial_\mu \sigma^C \partial_\nu \sigma^D = K_{CD} \frac{\partial \sigma^C}{\partial \chi^A} \frac{\partial \sigma^D}{\partial \chi^B} \partial_\mu \chi^A \partial_\nu \chi^B = G_{AB} \partial_\mu \chi^A \partial_\nu \chi^B,
\end{equation*}
where $G_{AB}$ is the new ``mass'' matrix.

In our particular case, we take
\begin{equation}
\label{sigmachi}
\sigma^1 = \sqrt{12 U_0} \tanh \frac{\chi^2}{\sqrt{6} M_\mathrm{Pl}}, \quad \sigma^2 = \frac{\sqrt{12 U_0}}{\cosh \frac{\chi^2}{\sqrt{6} M_\mathrm{Pl}}} \tanh \frac{\chi^1}{\sqrt{6} M_\mathrm{Pl}},
\end{equation}
and get the following matrix $G_{AB}$:
\begin{equation}
\begin{aligned}
G_{11} &= 1,\\
G_{12} &= G_{21} = 0,\\
G_{22} &= \cosh^2\left(\frac{\chi^1}{\sqrt{6} M_\mathrm{Pl}}\right).
\end{aligned}
\end{equation}

In terms of the new fields, the function $U$ takes the form
\begin{equation}
U(\chi^1, \chi^2) = \frac{U_0}{\cosh^2\left(\frac{\chi^1}{\sqrt{6} M_\mathrm{Pl}}\right)\cosh^2 \left(\frac{\chi^2}{\sqrt{6} M_\mathrm{Pl}}\right)}.
\end{equation}

Action (\ref{SE}) can be rewritten as
\begin{equation}
\label{einstein_action}
S_\mathrm{E} = \int d^4 x \sqrt{-g^{(\mathrm{E})}} \left(\frac{M^2_\mathrm{Pl}}{2} R^{(\mathrm{E})} - \frac{g^{(\mathrm{E})\mu\nu}}{2}\left[ \partial_\mu \chi^1 \partial_\nu \chi^1 +
\cosh^2\left(\frac{\chi^1}{\sqrt{6} M_\mathrm{Pl}}\right)  \partial_\mu \chi^2 \partial_\nu \chi^2\right] - V_\mathrm{E}\right),
\end{equation}
where
\begin{equation}
\label{VEchi}
V_\mathrm{E}=\frac{M^4_\mathrm{Pl}\left(\cosh\frac{\chi^1}{\sqrt{6} M_\mathrm{Pl}}\cosh\frac{\chi^2}{\sqrt{6} M_\mathrm{Pl}}\right)^2}{4U_0}\left(\frac{R_0}{2} +
144U_0\left[\tanh\frac{\chi^2}{\sqrt{6} M_\mathrm{Pl}}\right]^4 f\left(\frac{\tanh\frac{\chi^1}{\sqrt{6} M_\mathrm{Pl}}}{\sinh\frac{\chi^2}{\sqrt{6} M_\mathrm{Pl}}}\right)\right)\,.
\end{equation}

By varying action~(\ref{einstein_action}), we get the following equations:
\begin{equation}
\label{EquEinstein}
R^{(\mathrm{E})}_{\mu \nu} - \frac12 g^{(\mathrm{E})}_{\mu \nu} R^{(\mathrm{E})} = \frac{1}{M^2_\mathrm{Pl}} T^\mathrm{E}_{\mu \nu},
\end{equation}
\begin{equation}
\label{equchi1}
\Box^{(\mathrm{E})} \chi^1 - \frac{\sqrt{6}}{12M_\mathrm{Pl}}\sinh\left(\frac{\chi^1}{3\sqrt{6}M_\mathrm{Pl}}\right)  g^{(\mathrm{E})\mu\nu} \partial_\mu \chi^2 \partial_\nu \chi^2 - \frac{\partial V_\mathrm{E}}{\partial \chi^1} = 0;
\end{equation}
\begin{equation}
\label{equchi2}
\cosh^2\left(\frac{\chi^1}{\sqrt{6} M_\mathrm{Pl}}\right) \Box^{(\mathrm{E})} \chi^2 + \frac{\sqrt{6}}{6M_\mathrm{Pl}}\sinh\left(\frac{\chi^1}{3\sqrt{6}M_\mathrm{Pl}}\right)
 g^{(\mathrm{E})\mu \nu} \partial_\mu \chi^1 \partial_\nu \chi^2 - \frac{\partial V_\mathrm{E}}{\partial \chi^2} = 0,
\end{equation}
where
\begin{equation}
\begin{aligned}
T^\mathrm{E}_{\mu \nu} =& \partial_\mu \chi^1 \partial_\nu \chi^1 + \cosh^2\left(\frac{\chi^1}{\sqrt{6} M_\mathrm{Pl}}\right) \partial_\mu \chi^2 \partial_\nu \chi^2\\
- &g^{(\mathrm{E})}_{\mu \nu} \left(\frac12 g^\mathrm{(E)\alpha \beta} \partial_\alpha \chi^1 \partial_\beta \chi^1 + \frac12 \cosh^2\left(\frac{\chi^1}{\sqrt{6} M_\mathrm{Pl}}\right)g^{(\mathrm{E})\alpha \beta} \partial_\alpha \chi^2 \partial_\beta \chi^2 + V_\mathrm{E}(\chi^1, \chi^2)\right).
\end{aligned}
\end{equation}

The d'Alembert operator $\Box^{(\mathrm{E})}$ acting on a scalar is
\begin{equation}
\Box^{(\mathrm{E})}=\frac{1}{\sqrt{-g^{(\mathrm{E})}}}\partial_\mu\left(\sqrt{-g^{(\mathrm{E})}}g^{(\mathrm{E})\mu\nu}\partial_\nu\right).
\end{equation}

\section{Integrable cosmological models with nonminimal coupling}

\subsection{The FLRW metric with conformal time}

To get cosmological solutions in the analytic form, we consider the  FLRW metric with
conformal time
\begin{equation}
\label{Fried}
ds^2=a^2(\tau)\left(-d\tau^2+\frac{dr^2}{1-Kr^2}+r^2d\theta^2+r^2\sin^2(\theta)d\varphi^2\right),
\end{equation}
where $a(\tau)$ is the scale function. A positive curvature index  $K$  is associated with a closed universe, $K=0$ with a flat universe and a negative $K$ with an open one.

In this metric, Eqs.~(\ref{equRmunu}) and (\ref{equfield}) with $C_A=1$ have the following forms:
\begin{equation}
\label{Frequoc00}
6U\left[h^2+K\right]+6h\dot{U}=\frac12\left[\left({\dot{\sigma}^1}\right)^2+\left({\dot{\sigma}^2}\right)^2\right]+a^2V,
\end{equation}
\begin{equation}
\label{Frequocii}
U\left[2\dot{h}+h^2+K\right]+\ddot{U}+H\dot{U}=\frac12 a^2V-\frac14 \left[\left({\dot{\sigma}^1}\right)^2+\left({\dot{\sigma}^2}\right)^2\right],
\end{equation}
\begin{equation}
\ddot{\sigma}^A+2h\dot{\sigma}^A-6U_{,\sigma^A}a^2R+a^2V_{,\sigma^A} = 0,
\label{KG1}
\end{equation}
where $h=\dot{a}/a$, and ``dots'' denote derivatives with respect to the conformal time $\tau$.

The Ricci scalar is
\begin{equation}\label{RFr}
    R=\frac{6}{a^2}\left(\frac{\ddot{a}}{a}+K\right).
\end{equation}

Integrating equation $R=R_0$, namely,
\begin{equation}
\label{equRFr}
    \ddot{a}+Ka=\frac{R_0}{6}a^3,
\end{equation}
we obtain
\begin{equation}\label{IntR}
    \dot{a}^2+Ka^2=\frac{R_0}{12}a^4+C,
\end{equation}
where $C$ is an integration constant.

Equation~(\ref{IntR}) has the general solution in terms of the Jacobi elliptic function:
\begin{equation}
\label{atauelliptic}
\begin{split}
a(\tau) &= \frac{\sqrt{6C}}{\sqrt{3K+\sqrt{9K^2-3R_0C\,}}}\\
\times&\mathrm{sn}\left(\frac{1}{6}\sqrt{18K+6\sqrt{9K^2-3R_0C\,}}\left(\tau-\tau_0\right)\left|\frac{\sqrt{R_0C\left(6K^2-R_0C+2K\sqrt{9K^2-3R_0C\,}\right)}}{R_0C-6K^2-2K\sqrt{9K^2-3R_0C\,}}\right.\right)\!,
\end{split}
\end{equation}
if $R_0>0$. For $R_0=0$, we get
\begin{equation}
\label{atauR0}
a(\tau)=\left\{
\begin{aligned}
a_1\sin(\sqrt{K}\tau)+a_2\cos(\sqrt{K}\tau),\quad&\quad K>0;\\
a_1\tau+a_0,\qquad\qquad\qquad\quad\quad\quad\,\,\,&\quad K=0;\\
a_1\sinh(\sqrt{-K}\tau)+a_2\cosh(\sqrt{-K}\tau),&\quad K<0;\\
\end{aligned}
\right.
\end{equation}
where $a_0$, $a_1$, and $a_2$ are arbitrary constants.

To find solutions for scalar fields, we substitute $\sigma^A=y^A/a$ into Eq.~(\ref{KG1}) and obtain a system
\begin{equation}
\ddot{y}^A+Ky^A+ a^3  V_{,\phi^A}\left(\frac{y^1}{a},\frac{y^2}{a}\right) = 0.
\end{equation}

For $V$ given by~(\ref{V_choice}), we get the following system of two equations:
\begin{equation}
\label{y_system}
\ddot{y}^A+Ky^A + \frac{\partial \tilde{V}}{\partial y^A} = 0,
\end{equation}
where $A=1,2\,$ and
\begin{equation*}
\tilde{V}\left(y^1,y^2\right)=\left(y^1\right)^4f\left(\frac{y^2}{y^1}\right).
\end{equation*}

System (\ref{y_system})  is a Hamiltonian one with
\begin{equation}
\label{H}
    {\cal{H}}=\frac12\left(p^1\right)^2+\frac12\left(p^2\right)^2+\frac{K}{2}\left[\left(y^1\right)^2+\left(y^2\right)^2\right]  + \tilde{V},
\end{equation}
 where $p_i$ are momenta. It has the first integral
\begin{equation}
\label{intE}
\frac12\left(\dot{y}^1\right)^2+\frac12\left(\dot{y}^2\right)^2+\frac{K}{2}\left[\left(y^1\right)^2+\left(y^2\right)^2\right]  + \tilde{V}= E,
\end{equation}
where $E$ is an integration constant.

To analyze the cosmological evolution we consider equations in the cosmic time, which is defined by the equation $dt=a(\tau)d\tau$.
Using
\begin{equation}
h=aH,\qquad \dot{U}=a\frac{dU}{dt},\qquad \dot{y}^A=a^2\frac{d{\sigma}^A}{dt}+a^2H\sigma^A,
\end{equation}
where $H$ is the Hubble parameter, we get Eq.~(\ref{Frequoc00}) in the following form:
\begin{equation}
\label{Frequoc00H}
6U\left[H^2+\frac{K}{a^2}\right]+6H\frac{dU}{dt}=\frac12\left[\frac{d{\sigma}^1}{dt}\right]^2+\frac12\left[\frac{d{\sigma}^2}{dt}\right]^2+V.
\end{equation}

Using Eq.~(\ref{intE}), we transform Eq.~(\ref{Frequoc00H}) to the following equation:
\begin{equation}
 6U_0H^2+6U_0\frac{K}{a^2}=\frac12 U_0R_0+\frac{E}{a^4}.
 \label{Fried-closed1}
 \end{equation}

 Equation (\ref{Fried-closed1}) has the following solutions for $R_0>0$:
 \begin{equation}\label{sola}
    a^2(t)=\frac{36K^2U_0^2-2EU_0R_0+12KU_0\sqrt{R_0}\,{\rm e}^{\pm\sqrt{3R_0}\left(t-t_0\right)/3}+R_0\,{\rm e}^{\pm2\sqrt{3R_0}\left(t-t_0\right)/3}}{2U_0R_0^{3/2}{\rm e}^{\pm\sqrt{3R_0}\left(t-t_0\right)/3}}\,,
 \end{equation}
 where $t_0$ is a constant. So, the Hubble parameter is
\begin{equation}
\label{HtK}
    H(t)=\pm\,\frac{\sqrt{3R_0}\left(2EU_0R_0-36K^{2}U_0^{2}+R_0\,{\rm e}^{\pm\sqrt{3R_0}\left(t-t_0\right)/3} \right)}{6\left(36K^2U_0^{2}
    +12KU_0\sqrt{R_0}\,{\rm e}^{\pm\sqrt{3R_0}\left(t-t_0\right)/3}+R_0\,{\rm e}^{\pm2\sqrt{3R_0}\left(t-t_0\right)/3}-2U_0R_0E\right)}\,.
\end{equation}

In the case of $K=0$, the Hubble parameter has the following simple form:
\begin{equation}
H(t)=\frac{\sqrt{3R_0}\left(\mathrm{e}^{2\sqrt {3R_0}t/3}+B\right)}{6\left(\mathrm{e}^{2\,
\sqrt {3R_0}t/3}-B\right)}\,,
\end{equation}
where $B$ is a constant.

For $R_0=0$, one gets
\begin{equation}
a^2(t)=\left\{
\begin{split}
&\frac{E-6K^2U_0(t-t_0)^2}{6U_0K},\quad K\neq 0,\\
& \frac{\sqrt{6EU_0\,}}{3U_0}|t-t_0|\,,\quad K=0\,.
\end{split}
\right.
\end{equation}

The Hubble parameter is
\begin{equation}
H(t)=\left\{
\begin{split}
&\frac{6K^2U_0\left(t-t_0\right)}{6K^2U_0\left(t-t_0\right)^{2}-E},\quad K\neq 0\\
& \frac{1}{2(t-t_0)},\quad K=0\,.
\end{split}
\right.
\end{equation}

 Equation~(\ref{Fried-closed1}) coincides with the equation obtained in Ref.~\cite{Boisseau:2015hqa} for $K=0$ and in Ref.~\cite{Kamenshchik:2015cla} for an arbitrary $K$ in the case of the corresponding single-field model.
 The detail analysis of the Universe evolution in depending on the values of the constants $K$, $R_0$, and $E$, with conditions of the existence of bounce solutions, is given in Ref.~\cite{Kamenshchik:2015cla} without explicit expressions for $a(t)$ and $H(t)$. We do not repeat it in this paper, because it is also valid for the considered two-field models.

Note that the explicit form of the $a(\tau)$ does not depend on the form of the potential and is given by Eq.~(\ref{atauelliptic}) for $R_0>0$ and Eq.~(\ref{atauR0}) for $R_0=0$.
To get the general solutions in the conformal time, it is suitable to use expressions for $H(\tau)$ and $a(\tau)$ in explicitly real forms.

We present formulas for the case of $K=0$ and $R_0>0$:
\begin{enumerate}
\item If $E > 0$, then
\begin{equation}
H(\tau) = \sqrt{\frac{R_0}{3}}\frac{\mathrm{dn}\left(\left.a_0\sqrt{\frac{R_0}{3}}\left(\tau-\tau_0\right)\,\right|\frac12\right)}{1 - \mathrm{cn}\left(\left.a_0\sqrt{\frac{R_0}{3}}\left(\tau-\tau_0\right)\,\right|\frac12\right)},
\end{equation}
\begin{equation}
a(\tau) =  \frac{a_0\, \mathrm{sn}\left(\left.a_0\sqrt{\frac{R_0}{3}}\left(\tau-\tau_0\right)\,\right|\frac12\right)}{1+\mathrm{cn}\left(\left.a_0\sqrt{\frac{R_0}{3}}\left(\tau-\tau_0\right)\,\right|\frac12\right)},
\quad a_0 = \left(\frac{2E}{U_0 R_0}\right)^{\frac14},
\end{equation}
\begin{equation*}
\tau(t) = \tau_0 + \frac{1}{a_0}\sqrt{\frac{3}{R_0}}F\left(\arccos\left(\frac{a_0^2 - a^2}{a_0^2 + a^2}\right)\left|\frac12\right.\right),
\end{equation*}
\begin{equation*}
a_0\sqrt{\frac{R_0}{12}}\left(\tau-\tau_0\right) \in \left(0,  K\left(\frac12\right)\right).
\end{equation*}

\item  If $E < 0$, then
\begin{equation}
a(\tau) =  \frac{a_0}{\mathrm{cn}\left(\left.a_0\sqrt{\frac{R_0}{6}}\left(\tau-\tau_0\right)\,\right|\frac12\right)},\quad a_0 = \left(\frac{-2E}{U_0 R_0}\right)^{\frac14},
\end{equation}
\begin{equation}
H(\tau) = \sqrt{\frac{R_0}{6}}\,\mathrm{sn}\left(\left.a_0\sqrt{\frac{R_0}{6}}\left(\tau-\tau_0\right)\,\right|\frac12\right)\mathrm{dn}\left(\left.a_0\sqrt{\frac{R_0}{6}}\left(\tau-\tau_0\right)\,\right|\frac12\right),
\end{equation}
\begin{equation*}
\tau(t) = \tau_0 + \frac{1}{a_0}\sqrt{\frac{6}{R_0}}F\left(\arcsin\left(\sqrt{\frac{a^2 - a_0^2}{a^2}}\right)\left|\frac12\right.\right),
\end{equation*}
\begin{equation*}
a_0\sqrt{\frac{R_0}{6}}\left(\tau-\tau_0\right) \in \left(-K\left(\frac12\right), K\left(\frac12\right)\right).
\end{equation*}

\item If $E = 0$, then
\begin{equation}
H(\tau) = \pm\sqrt{\frac{R_0}{12}}, \qquad a(\tau) = \frac{a_0}{1 \mp a_0 \sqrt{R_0/12}\left(\tau-\tau_0\right)},
\end{equation}
\begin{equation*}
\tau(t) = \tau_0 \pm \frac{1}{a_0}\sqrt{\frac{12}{R_0}}\left(1 - e^{\mp \sqrt{R_0/12}(t-t_0)}\right) =\tau_0 \pm \frac{1}{a_0}\sqrt{\frac{12}{R_0}}\left(1 -\frac{a_0}{a}\right),
\end{equation*}
\begin{equation*}
\pm a_0\sqrt{\frac{R_0}{12}}\left(\tau-\tau_0\right) \in \left(-\infty, 1\right).
\end{equation*}
\end{enumerate}

Here $F$ is the incomplete elliptic integral of the first kind, defined as
\begin{equation}
F\left(\left.\varphi\,\right|m\right) = \int\limits_0^\varphi \frac{d\theta}{\sqrt{1 - m \sin^2 \theta}},
\end{equation}
$K(m) \equiv F(\pi / 2 \left| m\right.)$ is the complete elliptic integral of the first kind, and the Jacobi elliptic functions $\mathrm{sn}$, $\mathrm{cn}$, and $\mathrm{dn}$ have been used.

Solving Eq.~(\ref{Fried-closed1}) with an arbitrary $K$, one obtains:
\begin{enumerate}
\item at $\alpha^2 > 0$, $E / U_0 > 0$, $K > 0$:
\begin{equation}
a(\tau) = \sqrt{\alpha + \frac{6 K}{R_0}} \mathrm{dc}\left(\left.\sqrt{\alpha\frac{R_0}{12} + 2 K}(\tau - \tau_0)\right|\frac{6 K - \alpha R_0}{6 K + \alpha R_0}\right);
\end{equation}
\item at $\alpha^2 > 0$, $E / U_0 > 0$, $K < 0$:
\begin{equation}
a(\tau) = \sqrt{-\alpha + \frac{6 |K|}{R_0}} \mathrm{sc}\left(\left.\sqrt{\alpha\frac{R_0}{12} + 2 |K|}(\tau - \tau_0)\right|\frac{2\alpha R_0}{6 |K| + \alpha R_0}\right);
\end{equation}
\item at $\alpha^2 > 0$, $E / U_0 < 0$:
\begin{equation}
a(\tau) = \sqrt{\alpha + \frac{6 K}{R_0}}\mathrm{nc}\left(\left.\sqrt{\alpha\frac{R_0}{6}}(\tau - \tau_0)\right|\frac12 - \frac{3 K}{R_0 \alpha}\right);
\end{equation}
\item at $\alpha^2 < 0$:
\begin{equation}
a(\tau) = \left(\frac{2 E}{R_0 U_0}\right)^{1/4} \frac{\mathrm{sn}\left(\left.\sqrt{\frac{R_0}{3}}\left(\frac{2 E}{R_0 U_0}\right)^{1/4}(\tau - \tau_0)\right|\frac12 + \frac{3 K }{R_0}\sqrt{\frac{R_0 U_0}{2 E}}\right)}{1 + \mathrm{cn}\left(\left.\sqrt{\frac{R_0}{3}}\left(\frac{2 E}{R_0 U_0}\right)^{1/4}(\tau - \tau_0)\right|\frac12 + \frac{3 K }{R_0}\sqrt{\frac{R_0 U_0}{2 E}}\right)}.
\end{equation}
\end{enumerate}
 Here
\begin{equation*}
\alpha = \sqrt{\left(\frac{6 K}{R_0}\right)^2 - \frac{2 E}{R_0 U_0}}, \quad \mathrm{dc} = \frac{\mathrm{dn}}{\mathrm{cn}},\quad \mathrm{sc} = \frac{\mathrm{sn}}{\mathrm{cn}},\quad \mathrm{nc} = \frac{1}{\mathrm{cn}}.
\end{equation*}
In the two last cases $K$ is arbitrary.

Also, for the special case $E = 0$, $K = 0$, one has
\begin{equation}
a(\tau) = \frac{a_0}{1 \pm a_0 \sqrt{\frac{R_0}{12}}(\tau - \tau_0)}.
\end{equation}

\section{Potentials and solutions of the field equations}

\subsection{The way to get the general cosmological solution of the CCM}

In the Einstein frame, the FLRW metric (\ref{Fried}) becomes
\begin{equation}
\label{EinFreimMetric}
ds^2 =a_\mathrm{E}^2(\tau)\left( {}- d\tau^2 + \frac{dr^2}{1-Kr^2}+r^2d\theta^2+r^2\sin^2(\theta)d\varphi^2\right),
\end{equation}
where
\begin{equation}
a_\mathrm{E}(\tau) =  \sqrt{\frac{U}{U_0}}a(\tau).
\label{aJE}
\end{equation}

In the FLRW metric (\ref{EinFreimMetric}), Eqs.~(\ref{EquEinstein})--(\ref{equchi2}) have the following form:
\begin{equation}
3 M_\mathrm{Pl}^2 \left(h_\mathrm{E}^2+K\right) = \frac12 \left(\dot{\chi}^1\right)^2 + \frac12 G_{22} \left(\dot{\chi}^2\right)^2 +a^2_\mathrm{E} V_\mathrm{E},
\end{equation}
\begin{equation}
-M_\mathrm{Pl}^2 \left(2 \dot{h}_\mathrm{E} + h_\mathrm{E}^2+K\right) = \frac12 \left(\dot{\chi}^1\right)^2 + \frac12 G_{22} \left(\dot{\chi}^2\right)^2 - a^2_\mathrm{E} V_\mathrm{E},
\end{equation}
\begin{equation}
\ddot{\chi}^1 + 2 h_\mathrm{E} \dot{\chi}^1 - \frac12 \frac{dG_{22}}{d\chi^1} \left(\dot{\chi}^2\right)^2 + a^2_\mathrm{E} \frac{\partial V_\mathrm{E}}{\partial \chi^1} = 0,
\end{equation}
\begin{equation}
G_{22}\ddot{\chi}_2 + 2 h_\mathrm{E} G_{22}\dot{\chi}_2 + \frac{dG_{22}}{d\chi^1} \dot{\chi}_1 \dot{\chi}_2 + a^2_\mathrm{E}\frac{\partial V_\mathrm{E}}{\partial \chi^2} = 0.
\end{equation}
Here ``dots'' denote $d/d\tau$, and $h_\mathrm{E}\equiv \dot{a}_\mathrm{E}/a_\mathrm{E} = a_\mathrm{E} H_\mathrm{E}$.

Our goal is to get the general solutions of these equations for the potentials given by formula~(\ref{VEchi}).
If we know the general solutions for the corresponding model in the Jordan frame, or, in other words, if the  functions $a(\tau)$, $\sigma^1(\tau)$, and $\sigma^2(\tau)$ are known, then the scale function $a_\mathrm{E}$
is given by Eq.~(\ref{aJE}) and  the Hubble parameter
\begin{equation}
\label{HEJ}
 H_\mathrm{E}(\tau) = \frac{M_\mathrm{Pl}}{\sqrt{2 U(\sigma^1, \sigma^2)}} \left(H(\tau) + \frac{1}{2 a(\tau) U(\sigma^1, \sigma^2)} \frac{d U}{d \tau}\right).
\end{equation}
Here we use the fact that conformal times for both frames coincide (up to an additive constant, which can be set to be zero without loss of generality), since $dt / a = dt_\mathrm{E} / a_\mathrm{E}$.

To get $\chi^1(\tau)$ and $\chi^2(\tau)$, we use the relations
\begin{equation}
\label{chisigma}
\chi^1 = \sqrt{6} M_\mathrm{Pl} \,\mathrm{arctanh}\left(\frac{\sigma^2}{\sqrt{12 U_0 - \left(\sigma^1\right)^2}}\right), \quad \chi^2 = \sqrt{6} M_\mathrm{Pl}\, \mathrm{arctanh}\left(\frac{\sigma^1}{\sqrt{12 U_0\,}}\right).
\end{equation}

\subsection{Construction of integrable models}

To get an integrable CCM we choose a particular form of the potential $\tilde{V}$ in the Jordan frame, for which system~(\ref{y_system}) is integrable. This system is a Hamiltonian one and has an integral of motion~(\ref{intE}).
To prove the integrability of a such Hamiltonian model one should get the second independent integral of motion or show that the model passes the Painleve test (the description of the Painleve test and the use of it in cosmology see, for example, Refs.~\cite{Ablowitz:1980ca,PhysRevLett.49.1539,CONTE1989383,Miritzis:2000js,Paliathanasis:2016tch,Vernov:2002rw,Leon:2022dwd}). Integrable models with polynomial potentials $\tilde{V}$ have been found in Refs.~\cite{PhysRevLett.49.1539,Hietarinta:1983wh}. Moreover, the full list of polynomial potentials, for which the second invariant is a polynomial in momenta of order four or less has been found in Ref.~\cite{Hietarinta:1983wh}. This list consists of two sets of potentials $\tilde{V}$.  The first set includes potentials for which system~(\ref{y_system}) is integrable for $K=0$ only. These potentials are
\begin{equation}
\tilde{V}_a=V_0\left[\frac43\left(y^1\right)^4+\left(y^1\right)^2\left(y^2\right)^2+\frac16\left(y^2\right)^4\right],
\end{equation}
\begin{equation}
\tilde{V}_b=V_0\left[\frac43\left(y^1\right)^4+\left(y^1\right)^2\left(y^2\right)^2+\frac{1}{12}\left(y^2\right)^4\right],
\end{equation}
and the potentials obtained by reflection: $y^1\rightarrow y^2,\quad y^2\rightarrow y^1$. The corresponding second integrals are presented in Refs.~\cite{PhysRevLett.49.1539,Hietarinta:1983wh}.

The second set corresponds to integrable systems with an arbitrary $K$. We consider these potentials in the next subsections in detail.

\subsection{Potential $V=\frac12{R_0 U_0}$}

We start from the case of a constant potential
\begin{equation}\label{V0}
    V_0=\frac12{R_0 U_0}.
\end{equation}

The corresponding potential in the Einstein frame is
\begin{equation}
\label{VE0}
V_\mathrm{E}=\frac{M^4_\mathrm{Pl}R_0}{8U_0}\cosh^4\left(\frac{\chi^1}{\sqrt{6} M_\mathrm{Pl}}\right)\cosh^4\left(\frac{\chi^2}{\sqrt{6} M_\mathrm{Pl}}\right)\,.
\end{equation}

For $\tilde{V}\equiv 0$, Eq.~(\ref{y_system}) has the following solutions:
\begin{equation}
\label{yAVconst}
y^A=\left\{
\begin{aligned}
B_1^A\sin(\sqrt{K}\tau)+B_0^A\cos(\sqrt{K}\tau),\quad\qquad &\quad K>0;\\
B_1^A\tau+B_0^A,\qquad\qquad\qquad\quad\quad\quad\qquad\,\,\,&\quad K=0;\\
B_1^A\sinh(\sqrt{-K}\tau)+B^A_0\cosh(\sqrt{-K}\tau),\,\,&\quad K<0;\\
\end{aligned}
\right.
\end{equation}
where $B_0^A$ and $B_1^A$ are integration constants.

Using the expression (\ref{atauelliptic}), we get $\sigma^A(\tau)$. It allows us to get the general solution for the CCM with the potential (\ref{VE0}) by Eqs.~(\ref{aJE}) and (\ref{chisigma}).

\subsection{Potential $V=\frac12{R_0 U_0} + c_1 \left(\sigma^1\right)^4+ c_2 \left(\sigma^2\right)^4$}

Let us choose the potential $V$ as
\begin{equation}
\label{pot_sum}
V_1(\sigma^1, \sigma^2) = \frac{R_0 U_0}{2} + c_1 \left(\sigma^1\right)^4+ c_2 \left(\sigma^2\right)^4,
\end{equation}
where $c_1$ and $c_2$ are constant.

 System~(\ref{y_system}) with the potential (\ref{pot_sum}), is completely separable, and we have two independent integrals of motion:
\begin{equation}
\label{y_system_fin}
\frac12 \left(\frac{d y^A}{d \tau}\right)^2+\frac{K}{2}\left(y^A\right)^2 + c_A \left(y^A\right)^4= E_A,
\end{equation}
where $E_A$ are integration constants.

System~(\ref{y_system_fin}) includes two independent equations. Similar equations arise in the single-field model with hyperbolic potential, for which exact solutions with nonzero $K$ have been obtained in Ref.~\cite{Bars:2012mt}. In the appendix of Ref.~\cite{Bars:2012mt}, the authors list the complete set of analytic solutions to these equations in the various regions of the parameter space.

In this paper, we write out only solutions of Eq.~(\ref{y_system_fin})  with $K=0$; the form of these solutions depends on the signs of $c_A$ and $E_A$. Namely, we obtain:
\begin{enumerate}
\item $E_A > 0, \quad c_A > 0$:
\begin{equation}
y^A(\tau) = \left(\frac{E_A}{c_A}\right)^\frac14 \mathrm{cn}\left.\left(\pm 2\left(E_A c_A\right)^\frac14 (\tau - \tau_i) + \mathrm{cn}^{-1}\left(u_A\left|\frac12\right.\right)\right| \frac12\right);
\end{equation}
\item $E_A < 0, \quad c_A < 0$:
\begin{equation}
y^A(\tau) = \frac{\left(\frac{E_A}{c_A}\right)^\frac14}{\mathrm{cn}\left.\left(\pm 2\left(E_A c_A\right)^\frac14 (\tau - \tau_i) + \mathrm{cn}^{-1}\left.\left(\frac{1}{u_A}\right|\frac12\right)\right| \frac12\right)};
\end{equation}
\item $E_A > 0, \quad c_A < 0$:
\begin{equation}
y^A(\tau) = \left|\frac{E_A}{c_A}\right|^\frac14\frac{\mathrm{sn}\left(\pm\left.2\sqrt{2}\left|E_A c_A\right|^\frac14 (\tau - \tau_i) + \mathrm{sn}^{-1}\left(\left.\frac{2u_A}{1+u_A^2}\right|\frac12\right)\right| \frac12\right)}{1+\mathrm{cn}\left(\pm\left.2\sqrt{2}\left|E_A c_A\right|^\frac14 (\tau - \tau_i) + \mathrm{cn}^{-1}\left(\left.\frac{1-u_A^2}{1+u_A^2}\right|\frac12\right)\right| \frac12\right)}.
\end{equation}
\end{enumerate}
Here constants $u_A = |c_A/E_A|^{1/4} y^A(\tau_i)$ and the choice of sign in ``$\pm$'' are determined by initial conditions at $\tau=\tau_i$. Obviously, there is no solution when $E_A < 0$ and $c_A > 0$.

For $c_A = 0$, $E_A$ has to be non-negative, and we have $y^A(\tau)$ given by Eq.~(\ref{yAVconst}).

For the potential $V$ given by ~(\ref{pot_sum}), the corresponding potential $V_\mathrm{E}$ has the following form in terms of the new fields:
\begin{equation}
\begin{aligned}
\label{VE1}
V_\mathrm{E} &= \frac{M^4_\mathrm{Pl}}{4 U_0^2}\left[\frac12 R_0 U_0 \cosh^4\left(\frac{\chi^1}{\sqrt{6} M_\mathrm{Pl}}\right) \cosh^4\left(\frac{\chi^2}{\sqrt{6} M_\mathrm{Pl}}\right) \right.\\
&\left. + (12 U_0)^2 c_1 \cosh^4\left(\frac{\chi^1}{\sqrt{6} M_\mathrm{Pl}}\right) \sinh^4\left(\frac{\chi^2}{\sqrt{6} M_\mathrm{Pl}}\right) + (12 U_0)^2 c_2 \sinh^4\left(\frac{\chi^1}{\sqrt{6} M_\mathrm{Pl}}\right)\right].
\end{aligned}
\end{equation}

If we put $\chi^2=0$, then $V_\mathrm{E}$ coincides with the potential from the single-field model proposed by Bars and Chen in Ref.~\cite{Bars:2010zh}:
\begin{equation}
\label{VBC}
V_\mathrm{BC} = \frac{M^4_\mathrm{Pl}}{4 U_0^2}\left[\frac12 R_0 U_0 \cosh^4\left(\frac{\chi^1}{\sqrt{6} M_\mathrm{Pl}}\right) + (12 U_0)^2 c_2 \sinh^4\left(\frac{\chi^1}{\sqrt{6} M_\mathrm{Pl}}\right)\right].
\end{equation}

So, the proposed integrable model with potential (\ref{VE1}) is a two-field generalization of the known single-field integrable model~\cite{Bars:2010zh,Bars:2012mt}.

\subsection{Generalization of the potential $V=\frac12{R_0 U_0} + c_1 \left(\sigma^1\right)^4+ c_2 \left(\sigma^2\right)^4$}

The potential $V_1$ can be generalized. Indeed, let us consider the following potential:
\begin{equation}
\begin{split}
\tilde{V}_{1m}&=c_1\left(\cos(\theta_0)y^1+\sin(\theta_0)y^2\right)^4+c_2\left(\cos(\theta_0)y^2-\sin(\theta_0)y^1\right)^4\\
&=\left(c_1 \sin^{4}(\theta_0)+c_2 \cos^{4}(\theta_0)\right) \left(y^2\right)^{4}+4\cos(\theta_0)\sin(\theta_0)\left(c_1\sin^2(\theta_0)-c_2\cos^2(\theta_0) \right)
  {y^1} \left(y^2\right)^{3}\\
&{}+6\cos^{2}(\theta_0)\sin^{2}(\theta_0) \left( c_1+c_2 \right) \left(y^1\right)^2\left(y^2\right)^2\\
&{}+4\cos( \theta_0) \sin(\theta_0) \left( c_1 \cos^2(\theta_0)
-c_2 \sin^{2}(\theta_0)\right)\left(y^1\right)^3y^2 + \left( c_1\cos^{4}(\theta_0)+c_2\sin^{4}(\theta_0)\right)\left(y^1\right)^4,
\end{split}
\end{equation}
where $\theta_0$ is a constant.

After the rotation
\begin{equation}
\label{yturn}
\begin{split}
y^1&=\cos(\theta_0)z^1(\tau)-\sin(\theta_0)z^2(\tau),\\
y^2&=\sin(\theta_0)z^1(\tau)+\cos(\theta_0)z^2(\tau),
\end{split}
\end{equation}
we get
\begin{equation}
\label{pot_sumz}
\tilde{V}_{1}(z^1, z^2) = c_1 \left(z^1\right)^4+ c_2 \left(z^2\right)^4.
\end{equation}
Note that the kinetic part of the Hamiltonian~(\ref{H}) and the quadratic part of the potential are invariant under transformation~(\ref{yturn}), and so we obtain that the model with the potential
\begin{equation*}
V_{1m}(\sigma^1,\sigma^2)=\tilde{V}_{1m}(\sigma^1,\sigma^2)+\frac12 R_0U_0
\end{equation*}
 is integrable.

Using Eq.~(\ref{sigmachi}), we get the potential in the Einstein frame:
\begin{equation}
\label{VE1gen}
\begin{split}
    V_E&=36M^4_\mathrm{Pl}\cosh^{4}\left(\frac{\chi^1}{\sqrt{6}\,M_\mathrm{Pl}}\right)\left[\frac{R_0}{288\,U_0}\cosh^{4}\left(\frac{\chi^2}{\sqrt{6}\,M_\mathrm{Pl}}\right)\right. \\ &\left.{} + \left[c_1\cos^{4}(\theta_0)+c_2\sin^{4}(\theta_0)\right]
      \sinh^{4}\left(\frac{\chi^2}{\sqrt{6}\,M_\mathrm{Pl}}\right)\right. \\ &\left.{}
      +4\cos (\theta_0) \sin(\theta_0)\left[c_1\cos^{2}(\theta_0)-c_2\sin^{2}(\theta_0)\right]\tanh \left(\frac{\chi^1}{\sqrt{6}\,M_\mathrm{Pl}}\right)\sinh^{3}\left(\frac{\chi^2}{\sqrt{6}\,M_\mathrm{Pl}}\right)\right. \\ &\left.{}+6\left( c_1+c_2 \right)\cos^{2}(\theta_0)
        \sin^{2}(\theta_0)\tanh^{2} \left(\frac{\chi^1}{\sqrt{6}\,M_\mathrm{Pl}}\right)\sinh^{2}\left(\frac{\chi^2}{\sqrt{6}\,M_\mathrm{Pl}}\right)
            \right. \\&\left.{}-4\cos(\theta_0) \sin (\theta_0) \left[ c_2\cos^{2}(\theta_0)-c_1\sin^{2}(\theta_0)\right]\tanh^{3}\left(\frac{\chi^1}{\sqrt{6}\,M_\mathrm{Pl}}\right)\sinh\left(\frac{\chi^2}{\sqrt{6}\,M_\mathrm{Pl}} \right)
              \right. \\ &\left.{} +  \left[ c_1\sin^{4}(\theta_0)+c_2\cos^{4}(\theta_0) \right]\tanh^{4} \left(\frac{\chi^1}{\sqrt{6}\,M_\mathrm{Pl}}\right)  \right]\,.
\end{split}
\end{equation}

The general solution for the CCM with this potential can be found by using Eqs.~(\ref{aJE}) and (\ref{chisigma}).
Note that $\theta$, $c_1$, and $c_2$ are arbitrary parameters. So, we obtain new three-parametric sets of integrable models both in the Jordan and Einstein frames.

\subsection{Potential $V=\frac12{R_0 U_0} + c\left(\left(\sigma^1\right)^2 + \left(\sigma^2\right)^2\right)^2$}

For the potential
\begin{equation}
V_2=\frac12{R_0 U_0} + c\left(\left(\sigma^1\right)^2 + \left(\sigma^2\right)^2\right)^2,
\end{equation}
where $c$ is a nonzero constant, general solutions can also be explicitly written in terms of the Jacobi elliptic functions. In this case, Eq.~(\ref{intE}) takes the form
\begin{equation}
\frac12\left(\frac{d y^1}{d \tau}\right)^2 + \frac12 \left(\frac{d y^2}{d \tau}\right)^2 + \frac{K}{2}\left(\left(y^1\right)^2 + \left(y^2\right)^2\right) + c\left(\left(y^1\right)^2 + \left(y^2\right)^2\right)^2 = E.
\end{equation}
By introducing ``polar coordinates,''
\begin{equation}
y^1(\tau) = \rho(\tau) \cos \left(\theta(\tau)\right),\quad y^2(\tau) = \rho(\tau) \sin \left(\theta(\tau)\right),
\end{equation}
it can be rewritten as
\begin{equation}
\frac12\left(\frac{d \rho}{d \tau}\right)^2 + \frac12 \rho^2 \left(\frac{d \theta}{d \tau}\right)^2 + \frac{K}{2}\rho^2 + c \rho^4 = E.
\end{equation}

In analogy with classical mechanics, one may observe that $\theta$ is a cyclic coordinate, which means that the second integral is
\begin{equation}
\rho^2 \frac{d \theta}{d \tau} = L = \mathrm{const},
\end{equation}
and the equation for $\rho$ reads
\begin{equation}
\frac12\left(\frac{d \rho}{d \tau}\right)^2 + \frac{L^2}{2 \rho^2} + \frac{K}{2}\rho^2  + c \rho^4 = E.
\end{equation}

  So, we obtain that the model with potential $V_2$ is integrable for an arbitrary value of $K$.

In what follows, we restrict ourselves to the case of $c > 0$ only. In such a case the solution for $\rho$ reads
\begin{equation}
\rho(\tau) = \sqrt{\rho_0^2 - (\rho_0^2 - \rho_1^2)\, \mathrm{sn}^2\left(\left.\frac{\sqrt{v_0-v_2}}{2\beta}\tau + C_\rho \right| \frac{v_0-v_1}{v_0-v_2}\right)}.
\end{equation}
Here, $C_\rho$ is a constant of integration,
\begin{equation*}
\rho^2_{0, 1} = \alpha v_{0, 1},\quad v_k = \cos \left(\frac13 \arccos\left(-\frac{\frac{K^3}{216} + \frac{1}{12}K E c + \frac{1}{4}L^2 c^2}{\left(\frac13 E c + \frac{1}{12} K^2\right)^{3/2}} \right) - \frac{2 \pi k}{3}\right) - \frac{K}{2 c \alpha}, \quad k = 0, 1, 2,
\end{equation*}
\begin{equation*}
\alpha = 2\sqrt{\frac{E}{3 c} + \frac{K^2}{12 c^2}}, \quad \beta = \frac{1}{\sqrt{8 c \alpha}}.
\end{equation*}
The solution for $\theta$ is
\begin{equation}
\theta(\tau) =\frac{2 \beta L^2}{\alpha v_0 \sqrt{v_0 - v_2}}\Pi\left(\left.1-\frac{v_1}{v_0}; \frac{\sqrt{v_0-v_2}}{2\beta}\tau + C_\rho \right| \frac{v_0-v_1}{v_0-v_2}\right) + C_\theta.
\end{equation}
Here
\begin{equation*}
\Pi\left(\left.n; u\right|m\right) \equiv \int\limits_0^{u} \frac{d w}{1 - n\, \mathrm{sn}^2\left(\left.w\right|m\right)}
\end{equation*}
is the incomplete elliptic integral of the third kind, and $C_\theta$ is a constant of integration.

For the potential $V_2$, the corresponding potential $V_\mathrm{E}$ takes the form
\begin{equation}
\begin{aligned}
\label{VE2}
V_\mathrm{E} &= \frac{M^4_\mathrm{Pl}}{4 U_0^2}\left[\frac12 R_0 U_0 \cosh^4\left(\frac{\chi^1}{\sqrt{6} M_\mathrm{Pl}}\right) \cosh^4\left(\frac{\chi^2}{\sqrt{6} M_\mathrm{Pl}}\right) \right.\\
&\left. + (12 U_0)^2 c \left( \cosh^2\left(\frac{\chi^1}{\sqrt{6} M_\mathrm{Pl}}\right) \sinh^2\left(\frac{\chi^2}{\sqrt{6} M_\mathrm{Pl}}\right) + \sinh^2\left(\frac{\chi^1}{\sqrt{6} M_\mathrm{Pl}}\right)\right)^2\right].
\end{aligned}
\end{equation}

The general solution for the CCM with this potential can be found by using Eqs.~(\ref{aJE}) and (\ref{chisigma}).

\section{CONCLUSIONS}

In this paper, we have constructed new integrable chiral cosmological models with two scalar fields and potentials in terms of hyperbolic functions. These potentials are presented in Eqs.~(\ref{VE0}), (\ref{VE1}), (\ref{VE1gen}), and (\ref{VE2}).
The general solutions have been found in terms of the Jacobi elliptic functions of the conformal time.

To get these solutions we have used the corresponding models in the Jordan frame, for which the Ricci scalar is an integral of motion.
The integration of the initial system of equations reduces to the integration of system (\ref{y_system}) of two second-order equations, that has one integral of motion for any potential $\tilde{V}$ given by Eq.~(\ref{V_choice}).
This system is a Hamiltonian one. The existence of nonzero spatial curvature adds quadratic terms proportional to $K$ to the Hamiltonian~(\ref{H}).
In contrast to the single-field model, the existence of one integral of motion is not enough to obtain integrable models in the FLRW metric.
In this paper, we construct new examples of the potential $\tilde{V}$, for which system (\ref{y_system}) has two independent integrals of motion for an arbitrary $K$ and, hence, the corresponding two-field CCM is integrable not only in the spatially flat FLRW metric, but also in the open and closed FLRW metric. All integrable cases of the Hamiltonian (\ref{H}) with fourth-order polynomial potentials, for which the second invariant is a polynomial in momenta of order four or less, have been found in Ref.~\cite{Hietarinta:1983wh}. To the best of our knowledge the existence of an integrable system (\ref{y_system}) with a more complicated potential, for example, a nonpolynomial one, is an open question.

One of the important questions of the theoretical cosmology has to do with a possible evolution of the Universe before inflation. In particular, the question of the existence of a smooth connection between gravity and antigravity~\cite{Bars:2011aa,Bars:2013qna,Wetterich:2014zta,Kamenshchik:2016gcy} and the consideration of a cyclic universe~\cite{Steinhardt:2001st,Khoury:2003rt,Bars:2011mh,Lehners:2008vx,Ellis:2015bag} attract a lot of attention.
Single-field integrable cosmological models are actively used in such investigations. Moreover, integrable models can be used as main parts for more realistic models of gravity that confirm with observational data.
For example, the model proposed in Ref.~\cite{Bars:2011aa} can describe the Universe before and at the beginning of inflation, and so exact solutions can be used to calculate inflationary parameters analytically, without resorting to the slow-roll approximation. At the same time, some new terms should be added to get the exit from inflation. A similar situation arises with the integrable model proposed in Ref.~\cite{Boisseau:2015hqa}, with the goal to describe the possibility of producing a bouncing universe in the framework of scalar-tensor gravity. In this model, the Hubble parameter is a monotonically increasing function, and to get more realistic models with a nonmonotonic Hubble parameter one needs to slightly modify the specified integrable model~\cite{Boisseau:2016pfh,Pozdeeva:2016cja}.

Two-field CCMs are actively used in cosmology to describe the evolution of the observable Universe, including inflation, primordial black hole formation and dark energy~\cite{Elizalde:2004mq,Chervon:2013btx,Karananas:2016kyt,Gorbunov:2018llf,Christodoulidis:2019mkj,Gundhi:2020kzm,Braglia:2020eai,He:2020ivk,Geller:2022nkr,Braglia:2022phb,Cado:2023zbm,Pozdeeva:2024hfq}. The constructed models can be considered as two-field generalizations of the single-field integrable model proposed in Ref.~\cite{Bars:2010zh}, which has been actively investigated as a possible description of the evolution of the early Universe.
We hope that the proposed two-field integrable models will be useful for describing the Universe evolution.

\section*{ACKNOWLEDGEMENTS}

V. R. I. is supported by the ``BASIS'' Foundation Grant No. 22-2-2-6-1.

\section*{REFERENCES}

\bibliographystyle{apsrev}
\bibliography{TwoFieldModel}

\end{document}